\documentclass[10pt, conference, letterpaper]{IEEEtran}
\usepackage{amsmath,amssymb,amsfonts,amsthm,romannum}
\usepackage{algorithm,algorithmic}
\usepackage{graphicx,xcolor}
\usepackage[caption=false]{subfig}
\usepackage{balance}
\graphicspath{{Figure/}}
\def\ie{{\it i.e.,\ \/}}
\def\eg{{\it e.g.,\ \/}}
\def\st{{\it s.t.,\ \/}}
\DeclareMathOperator{\blkdiag}{blkdiag}
\theoremstyle{definition}
\newtheorem{theorem}{Theorem}
\newtheorem{remark}{Remark}
\newtheorem{lemma}{Lemma}
\newtheorem{corollary}{Corollary}
\newtheorem{assumption}{Assumption}
\makeatletter
\def\blfootnote{\gdef\@thefnmark{}\@footnotetext}
\makeatother
\usepackage{flushend}
\begin{document}

\title
{Delay-Tolerant Constrained OCO with Application to Network Resource
Allocation
\author{
\IEEEauthorblockN{Juncheng Wang\IEEEauthorrefmark{1},
Ben Liang\IEEEauthorrefmark{1},
Min Dong\IEEEauthorrefmark{2},
Gary Boudreau\IEEEauthorrefmark{3}, and
Hatem Abou-zeid\IEEEauthorrefmark{3}}
\IEEEauthorblockA{\IEEEauthorrefmark{1}Department of Electrical and Computer
Engineering, University of Toronto, Canada,\\
\IEEEauthorrefmark{2}Department of Electrical, Computer and Software Engineering,
Ontario Tech University, Canada, \IEEEauthorrefmark{3}Ericsson
Canada, Canada\\}}}
\maketitle

\begin{abstract}
We consider online convex optimization (OCO) with multi-slot feedback delay, where an agent makes a sequence of online decisions to minimize the accumulation of time-varying convex loss functions, subject to short-term and long-term
constraints that are possibly time-varying. The current convex loss function and the long-term constraint function are revealed to the agent only after the decision is made, and they may be delayed for multiple time slots. Existing work on OCO under this general setting has focused on the static regret, which measures the gap of losses between the online decision sequence and an offline benchmark that is fixed over time. In this work, we consider both the static regret and the more practically meaningful dynamic regret, where the benchmark is a time-varying sequence of per-slot optimizers. We propose an efficient algorithm, termed Delay-Tolerant Constrained-OCO (DTC-OCO), which uses a novel constraint penalty with double regularization to tackle the asynchrony between information feedback and decision updates. We derive upper bounds on its dynamic regret, static regret, and constraint violation, proving them to be sublinear under mild conditions. We further apply DTC-OCO to a general network resource allocation problem, which arises in many systems such as data networks and cloud computing. Simulation results demonstrate substantial performance gain of DTC-OCO over the known best alternative.\blfootnote{This
work has been funded in part by Ericsson Canada and by the Natural
Sciences and Engineering Research Council (NSERC) of Canada.}
\end{abstract}

\section{Introduction}
\label{Sec:Introduction}

\begin{table*}[ht]
\renewcommand{\arraystretch}{1.1}
\vspace{-2mm}
\caption{Summary of Related Works on OCO}
\vspace{-2mm}
\label{tab:comp}
\centering
\resizebox{\textwidth}{!}{
\begin{tabular}{|c|c|c|c|c|c|c|c|c|c|c|c|c|c|c|c|c|c|}\hline
~ &\cite{OLBK}&\cite{introOCO}&\cite{Zinkevich}&\cite{OT}&\cite{E.C.Hall}&\cite{CDC}&\cite{Trade}&\cite{LTC-Toff}&\cite{LTC-HY}&\cite{Yu-SC}&\cite{T.Chen}&\cite{X.Cao}&\cite{Slow}&\cite{AD}&\cite{McMahan}&\cite{X.CaoTau}&DTC-OCO\\\hline\hline
Long-term time-invariant constraints    &~&~&~&~&~&~&Y&Y&Y&~&~&~&~&~&~&~&Y\\\hline
Long-term time-varying constraints      &~&~&~&~&~&~&~&~&~&Y&Y&Y&~&~&~&Y&Y\\\hline
Multi-slot feedback delay               &~&~&~&~&~&~&~&~&~&~&~&~&Y&Y&Y&Y&Y\\\hline
Static performance benchmark            &Y&Y&Y&Y&~&~&Y&Y&Y&Y&~&~&Y&Y&Y&Y&Y\\\hline
Dynamic performance benchmark           &~&~&Y&~&Y&Y&~&~&~&~&Y&Y&~&~&~&~&Y\\\hline
\end{tabular}}
\vspace{-5mm}
\end{table*}

Online convex optimization (OCO) has emerged as a promising solution to many machine learning, signal processing, and resource allocation problems in the presence of uncertainty \cite{OLBK}, \cite{introOCO}. Under the standard OCO setup, at the beginning of each time slot, an agent makes a decision from a known convex set. At the end of each slot, the system reveals information of the current convex loss function to the agent, and then the agent's loss is realized.  Due to the lack of in-time information of the current convex loss function, it is impossible for an agent to make an optimal decision at each slot. Instead, the agent aims at minimizing the \emph{regret}, \ie the performance gap between the online decision sequence and some performance benchmark.

Most of the early works on OCO studied the \emph{static} regret, which compares the online decision sequence with a static offline benchmark that has apriori information of all the convex loss functions. In the seminal work of OCO \cite{Zinkevich}, an online projected gradient descent algorithm was shown to achieve $\mathcal{O}(T^{\frac{1}{2}})$ static regret, where $T$ is the time horizon. The static regret was further reduced to $\mathcal{O}(\log{T})$  in \cite{OT} for strongly convex loss functions. However, when the optimum of the underlying system is time-varying, the static offline benchmark itself may perform poorly and achieving sublinear static regret may not be meaningful anymore. In \cite{Zinkevich}, a more useful \emph{dynamic} regret was also presented, where the benchmark was a time-varying solution sequence. This dynamic benchmark has been recognized as a more attractive but harder-to-track performance benchmark for OCO \cite{E.C.Hall}, \cite{CDC}. 

The above works all focused on OCO with short-term constraints that must be strictly satisfied. Subsequently, OCO with \emph{long-term} constraints was considered in \cite{Trade}. In this case, a desired OCO algorithm should provide sublinear \emph{constraint violation}, so that the time-averaged violation of each long-term constraint tends to zero as time approaches infinity. A trade-off between the regret and constraint violation bounds was shown in \cite{LTC-Toff}, and the constraint violation bound was further improved in \cite{LTC-HY}. These works assumed that the long-term constraints are time-invariant and known in advance, while recent works \cite{Yu-SC}\nocite{T.Chen}-\cite{X.Cao} studied OCO with \emph{time-varying} long-term constraints.

In practical systems, the decision maker often gains access to the system information only after some delay. For example, in wireless communications, data transmission relies on channel state information, which is usually delayed for multiple transmission frames after channel estimation, quantization, and feedback. In mobile computing, feedback delay can be caused by offloading tasks from remote devices and communicating through wireless channels. Under standard OCO, the decision maker receives information on the current loss function (and if applicable the long-term constraint functions) at the end of each slot when the decision is made, \ie the feedback information is delayed for only one slot. This is the model used in, \eg \cite{Zinkevich}\nocite{OT}\nocite{E.C.Hall}\nocite{CDC}\nocite{Trade}\nocite{LTC-Toff}\nocite{LTC-HY}\nocite{Yu-SC}\nocite{T.Chen}-\cite{X.Cao}, but it can be too restrictive for many practical applications. Therefore, \cite{Slow} initiated a study on OCO with multi-slot feedback delay. Additional delay-adaptive OCO algorithms were proposed in \cite{AD} and \cite{McMahan}, while a static regret bound was derived in recent work \cite{X.CaoTau} for time-varying constrained OCO with multi-slot feedback delay.

However, to the best of our knowledge, all existing works on OCO with multi-slot feedback delay have focused on the static regret, which may not be a meaningful performance metric for some inherently time-varying systems. In fact, the gap between the static and dynamic regrets can be as large as $\mathcal{O}(T)$ \cite{Gap}. In the presence of multi-slot feedback delay, whether sublinear dynamic regret is achievable for OCO is an open problem. The need to consider (time-varying and time-invariant) long-term  constraints further adds to this challenge. In this context, the main contributions of this paper are as follows:
\begin{itemize}

\item 
We propose an efficient algorithm, termed Delay-Tolerant Constrained-OCO (DTC-OCO), for OCO with multi-slot feedback delay and both short-term and long-term constraints. While all existing OCO algorithms use either no regularization or single regularization for decision updates, in DTC-OCO, we propose a novel constraint penalty with  double regularization to tackle the asynchrony between information feedback and decision updates, which improves tolerance to multi-slot delay and facilitates the performance bounding of DTC-OCO. 

\item 
We analyze the special structure of DTC-OCO due to double regularization and establish that it provides $\mathcal{O}(\max\{\tau^{\frac{1}{2}}T^{\frac{1+\delta}{2}},T^\nu\})$ dynamic regret and $\mathcal{O}(\max\{\tau^{\frac{1}{2}}T^{\frac{1}{2}},T^\nu\})$
static regret, where $\tau$ is the feedback delay, and $\delta$ and $\nu$ respectively represent the growth rates of the accumulated variation of a dynamic benchmark and constraint functions.
\item 
We derive $\mathcal{O}(\max\{T^{\frac{1-\delta}{2}},\tau{T^\kappa}\})$ constraint violation bound for DTC-OCO, where $\kappa$ represents the growth rate of the accumulated squared variation of constraint functions. In the special case of time-invariant constraints, even with multi-slot feedback delay, DTC-OCO recovers the known best constraint violation bound for constrained OCO with one-slot feedback delay.

\item
We apply DTC-OCO to a general network resource allocation problem, which arises in many systems such as data networks and cloud computing. Simulation  based on an example cloud computing system demonstrates that DTC-OCO is tolerant to feedback delay and substantially outperforms the known best alternative from \cite{X.CaoTau}.

\end{itemize}

The rest of this paper is organized as follows. In Section \ref{Sec:Related Work}, we present the related work. Section \ref{Sec:Problem Formulation and Online Algorithm} describes the problem formulation and performance metrics. We present DTC-OCO, provide its performance bounds, and discuss its  performance merits in Section \ref{Sec:Performance Bounds}. The application of DTC-OCO to network resource allocation is presented in Section \ref{Sec:Application to Network Resource Allocation}, followed by concluding remarks in Section \ref{Sec:Conclusions}.

\section{Related Work}
\label{Sec:Related Work}

In this section, we survey existing works on OCO with either long-term constraints or multi-slot feedback delay. The differences between these works and our work are summarized in Table.~\ref{tab:comp}.

Among existing works on OCO with long-term constraints, a saddle-point-typed algorithm was proposed in \cite{Trade} and achieved $\mathcal{O}(T^{\frac{1}{2}})$ static regret and $\mathcal{O}(T^{\frac{3}{4}})$ constraint violation. A follow-up work \cite{LTC-Toff}  provided $\mathcal{O}(T^{\max\{\mu,1-\mu\}})$ static regret and $\mathcal{O}(T^{1-\frac{\mu}{2}})$ constraint violation, where $\mu\in(0,1)$ is some trade-off parameter. A virtual-queue-based algorithm was proposed in \cite{LTC-HY} and established the known best $\mathcal{O}(1)$ constraint violation. Another virtual-queue-based algorithm was proposed in \cite{Yu-SC} for independent and identically distributed (i.i.d.) constraints. The saddle-point-typed and virtual-queue-based algorithms were respectively modified in \cite{T.Chen} and \cite{X.Cao} to deal with time-varying constraints. However, the above works were all confined to the standard OCO setting where the feedback information is delayed for only one slot.

Most existing works on OCO with multi-slot feedback delay focused on online problems with short-term constraints \cite{Slow}\nocite{AD}-\cite{McMahan}. The standard online gradient descent algorithm \cite{Zinkevich} was extended  in \cite{Slow} to provide $\mathcal{O}(\tau^{\frac{1}{2}}T^{\frac{1}{2}})$ static regret. Delay-adaptive online gradient descent algorithms were proposed in \cite{AD} and \cite{McMahan} to accommodate adversarial feedback delay. The impact of multi-slot feedback delay on OCO with long-term constraints was considered only in \cite{X.CaoTau}. However, \cite{X.CaoTau} only studied the static regret, which may not be an attractive performance metric for an inherently time-varying system. Furthermore, the constraint violation bound provided in \cite{X.CaoTau} was no less than $\mathcal{O}(T^{\frac{3}{4}})$ even for time-invariant constraints, inherited from the saddle-point-typed algorithm \cite{Trade}. Thus, the impact of multi-slot feedback delay on constrained OCO is still not well understood. Different from \cite{X.CaoTau}, we propose a novel constraint penalty with double regularization to achieve sublinear dynamic and static regret bounds, as well as a stronger constraint violation bound.

Constrained OCO is related to Lyapunov optimization \cite{Neely}, which makes use of the system state and queueing information to implicitly learn and adapt to changes in the system with unknown statistics. However, under this framework, the system states are commonly assumed to be i.i.d. or Markovian, while OCO solutions do not have such restriction. Furthermore, standard Lyapunov optimization relies on the current and accurate system state for decision updates. In the presence of feedback delay on the system state, one can apply Lyapunov optimization by leveraging historical information to predict the current system state with some errors \cite{Lotfinezhad10}. However, this way of dealing with feedback delay is equivalent to extending the standard Lyapunov optimization to inaccurate system states in \cite{H.Yu} and \cite{Wang20}. In this case, the optimality gap would be $\mathcal{O}(\sigma{T})$ with $\sigma$ being some measure of system inaccuracy, so that such an approach cannot lead to the sublinear dynamic regret bound that we seek.

In this work, we focus on centralized OCO. Distributed OCO is beyond our scope, and we refer interested readers to \cite{Duchi12}\nocite{Akbari17}\nocite{Jadbabaie18}\nocite{Nima20}-\cite{XYi20}.

\section{Constrained OCO with Multi-slot Delay}
\label{Sec:Problem Formulation and Online Algorithm}

\subsection{Problem Formulation}
\label{Subsec:Problem Formulation}

We consider a time-slotted system with time indexed by $t$. Let $\{f_t(\mathbf{x})\}$, where $f_t(\mathbf{x}):\mathbb{R}^n\to\mathbb{R}$, be a sequence of convex loss functions. Let $\{\mathbf{g}_t(\mathbf{x})\}$ be a sequence of convex long-term constraint functions, where $\mathbf{g}_t(\mathbf{x})=[g_t^1(\mathbf{x}),\dots,g_t^C(\mathbf{x})]^T:\mathbb{R}^n\to\mathbb{R}^C$ with $C$ being the number of constraints. Let $\mathcal{C}=\{1,\dots,C\}$. The loss function $f_t(\mathbf{x})$ and the constraint function $\mathbf{g}_t(\mathbf{x})$ can vary over time. We further consider short-term constraints represented by a compact convex set $\mathcal{X}_0\subseteq\mathbb{R}^n$. 
The goal of constrained OCO is to select a sequence of decisions $\{\mathbf{x}_t\}$ from $\mathcal{X}_0$ that minimizes the accumulated loss while ensuring that the long-term constraints are satisfied \cite{T.Chen}, \cite{X.Cao}. This leads to the following optimization problem:
\begin{align}
        \textbf{P1}:\quad\min_{\{\mathbf{x}_t\in\mathcal{X}_0\}} \quad &\sum_{t=1}^{T}f_t(\mathbf{x}_t)\notag\\
        \text{s.t.}\qquad&\sum_{t=1}^{T}\mathbf{g}_t(\mathbf{x}_t)\preceq\mathbf{0}.\label{eq:ltc}
\end{align}
Note that if the constraint functions are time-invariant, \ie $\mathbf{g}_t(\mathbf{x})=\mathbf{g}(\mathbf{x}),\forall{t}$, then $\textbf{P1}$ is simplified to the time-invariant constrained OCO problem considered in \cite{Trade}\nocite{LTC-Toff}-\cite{LTC-HY}. 

Under the standard constrained OCO setting \cite{Trade}\nocite{LTC-Toff}\nocite{LTC-HY}\nocite{Yu-Sc}\nocite{T.Chen}-\cite{X.Cao}, feedback information on the loss function $f_t(\mathbf{x})$ and the long-term constraint function $\mathbf{g}_t(\mathbf{x})$ is assumed to be delayed for only one slot, when it can be leveraged to make the new decision~$\mathbf{x}_{t+1}$.\footnote{We note that, to have a well-posed problem, information on the short-term constraints must be current. Furthermore, obviously delay is irrelevant to time-invariant long-term constraints \cite{Trade}\nocite{LTC-Toff}-\cite{LTC-HY}.} However, in many practical applications, \eg wireless transmission and mobile computing mentioned in Section~\ref{Sec:Introduction}, this assumption is rarely satisfied since the feedback information can be severely delayed. Therefore, in this work, we consider a general scenario where the feedback information of $f_t(\mathbf{x})$ and $\mathbf{g}_t(\mathbf{x})$ is delayed for $\tau\ge1$ slots, so that the decision maker receives it at the end of slot $t+\tau-1$ \cite{Slow}\nocite{AD}\nocite{McMahan}-\cite{X.CaoTau}. Different from similar works with only short-term constraints \cite{Slow}\nocite{AD}-\cite{McMahan}, the additional long-term constraints in (\ref{eq:ltc}) of $\textbf{P1}$ lead to a more complicated online problem, especially since the underlying system varies over time while the feedback information is delayed.

\subsection{Performance Metrics}
\label{Subsec:Performance Metrics}

Due to the lack of in-time information of the current loss and constraint functions under the OCO setting, it is impossible to obtain an optimal solution to $\textbf{P1}$.\footnote{In fact, even for the most basic OCO problem \cite{Zinkevich}
(\ie without long-term constraints (\ref{eq:ltc})), an optimal solution cannot be found \cite{OT}.}  Instead, a time-varying constrained OCO algorithm aims at selecting an online solution sequence $\{\mathbf{x}_t\}$ that is asymptotically no worse than some performance benchmark.

One common offline benchmark is
\begin{align}
        \mathbf{x}^\star\in\arg\min_{\mathbf{x}\in\mathcal{X}_0}\left\{\sum_{t=1}^Tf_t(\mathbf{x})|\mathbf{g}_t(\mathbf{x})\preceq\mathbf{0},\forall{t}\right\}\label{eq:xstar}
\end{align}
which is computed assuming all information of $\{f_t(\mathbf{x})\}$ and $\{\mathbf{g}_t(\mathbf{x})\}$ is known in advance. The performance gap between $\{\mathbf{x}_t\}$ and  $\mathbf{x}^\star$ is termed the static regret:
\begin{align}
        \text{RE}_{\text{s}}(T)\triangleq\sum_{t=1}^T[f_t(\mathbf{x}_t)-f_t(\mathbf{x}^\star)].\label{eq:Sreg}
\end{align}
For example, this static regret was used in \cite{X.CaoTau}, while \cite{Trade}\nocite{LTC-Toff}-\cite{LTC-HY} used a special case of it with time-invariant constraints.\footnote{Even with one-slot delay, \cite{S.Mannor09} showed that it is impossible to achieve sublinear static regret w.r.t. $\mathbf{x}^\circ\!\in\!\arg\min_{\mathbf{x}\in\mathcal{X}_0}\{\sum_{t=1}^Tf_t(\mathbf{x})|\sum_{t=1}^T\mathbf{g}_t(\mathbf{x})\!\preceq\!\mathbf{0}\}$
and sublinear constraint violations simultaneously. }  However, as a rather coarse performance metric, the static regret may not be a strong indicator for the actual algorithm performance especially when the underlying system is inherently time-varying.

A more attractive performance benchmark for time-varying constrained OCO is the dynamic benchmark $\{\mathbf{x}_t^\star\}$,  given by
\begin{align}
        \mathbf{x}_t^{\star}\in\arg\min_{\mathbf{x}\in\mathcal{X}_0} \{f_t(\mathbf{x})|\mathbf{g}_t(\mathbf{x})\preceq\mathbf{0}\}\label{eq:xtstar}
\end{align}
which is computed using the in-time information of $f_t(\mathbf{x})$ and $\mathbf{g}_t(\mathbf{x})$ at each slot $t$. The dynamic benchmark was originally proposed for OCO with short-term constraints \cite{Zinkevich} and was modified in \cite{T.Chen} and \cite{X.Cao} to incorporate long-term constraints. The corresponding dynamic regret is defined  as
\begin{align}
        \text{RE}_{\text{d}}(T)\triangleq\sum_{t=1}^{T}[f_t(\mathbf{x}_t)-f_t(\mathbf{x}_t^{\star})].\label{eq:Dreg}
\end{align}
In some cases, the gap between $\text{RE}_{\text{s}}(T)$ and $\text{RE}_{\text{d}}(T)$ can be as large as $\mathcal{O}(T)$ \cite{Gap}. In this paper, we provide upper bounds on both  $\text{RE}_{\text{s}}(T)$ and $\text{RE}_{\text{d}}(T)$ for a comprehensive performance study. 

To measure the accumulated violation of the long-term constraints, the constraint violation\footnote{The constraint violation is referred to as dynamic fit $\text{Fit}(T)\triangleq\Vert[\sum_{t=1}^T\mathbf{g}_t(\mathbf{x}_t)]^+\Vert_2$
in \cite{T.Chen}, where $[\mathbf{x}]^+\triangleq\max\{\mathbf{x},\mathbf{0}\}$ is the entry-wise positive projection operator. One can easily verify that the sublinearity of $\text{VO}^c(T),\forall{c}\in\mathcal{C}$ implies $\text{Fit}(T)$ being sublinear, and vice versa.} is defined for any $c\in\mathcal{C}$ as \cite{X.Cao}, \cite{X.CaoTau}
\begin{align}
        \text{VO}^c(T)\triangleq\sum_{t=1}^{T}g_t^c(\mathbf{x}_t).\label{eq:Vio}
\end{align}
Note that the constraint violation defined in \cite{Trade}\nocite{LTC-Toff}-\cite{LTC-HY} for time-invariant constraint $\mathbf{g}(\mathbf{x})$ is a special case of (\ref{eq:Vio}). Our study accommodates both time-varying and time-invariant constraints.

A constrained OCO algorithm is desired to provide both sublinear regrets, \ie $\text{RE}_{\text{d}}(T)=\mathbf{o}(T)$ and $\text{RE}_{\text{s}}(T)=\mathbf{o}(T)$, and sublinear constraint violation, \ie $\text{VO}^c(T)=\mathbf{o}(T)$. Sublinearity is important since it implies that the online decision is asymptotically no worse than the benchmark in terms of its time-averaged performance, and the long-term constraints are satisfied in the time-averaged sense. Unfortunately, with multi-slot feedback delay, existing literature only achieves sublinear static regret and $\mathcal{O}(T^\frac{3}{4})$ constraint violation \cite{X.CaoTau}. In this work, we propose an efficient algorithm DTC-OCO to provide sublinear dynamic and static regrets and a stronger constraint violation bound. 

\section{Delay-Tolerant Constrained Online Convex Optimization (DTC-OCO)}
\label{Sec:Performance Bounds}

In this section, we present details of DTC-OCO and study the impact of multi-slot feedback delay on the performance guarantees of DTC-OCO to provide regret and constraint violation bounds. We further give sufficient conditions under which DTC-OCO yields sublinear regret and constraint violation. Finally, the performance merits of DTC-OCO over existing constrained OCO algorithms are discussed.

\subsection{DTC-OCO Algorithm}
\label{Subsec:Delay-Tolerant Constrained OCO Algorithm}

DTC-OCO introduces a novel virtual queue vector $\mathbf{Q}_t=[Q_t^1,\dots,Q_t^C]^T$ for the long-term constraints (\ref{eq:ltc}), with the following updating rule for any $c\in\mathcal{C}$:
\begin{align}
        Q_{t}^c=\max\left\{-\gamma g_{t-\tau}^c(\mathbf{x}_{t}),Q_{t-1}^{c}+\gamma{g}_{t-\tau}^c(\mathbf{x}_{t})\right\}\label{eq:VQ}
\end{align}
where $\gamma>0$ is a step-size parameter. The role of $\mathbf{Q}_t$ is similar to a Lagrangian multiplier vector for $\textbf{P1}$ or a backlog queue for the constraint violation, which are concepts used in \cite{LTC-HY}, \cite{Yu-SC}, and \cite{X.Cao}. However, unique to our proposed approach, $g_{t-\tau}^c(\mathbf{x}_t)$ is the $\tau$\textit{-slot delayed} constraint violation caused by the \textit{current} decision, and it needs to be scaled by an appropriate $\gamma$ factor. We then convert $\textbf{P1}$ to solving a per-slot problem at each slot $t>\tau$ with short-term constraints only, given by
\begin{align*}
        \textbf{P2}:~\min_{\mathbf{x}\in\mathcal{X}_0}\quad&[\nabla f_{t-\tau}(\mathbf{x}_{t-\tau})]^T(\mathbf{x}-\mathbf{x}_{t-\tau})\\
        &+[\mathbf{Q}_{t-1}+\gamma\mathbf{g}_{t-\tau-1}(\mathbf{x}_{t-1})]^T[\gamma\mathbf{g}_{t-\tau}(\mathbf{x})]\\
        &+\alpha\Vert\mathbf{x}-\mathbf{x}_{t-\tau}\Vert_2^2+\eta\Vert\mathbf{x}-\mathbf{x}_{t-1}\Vert_2^{2}
\end{align*}
where $\alpha,\eta>0$ are two step-size parameters.

DTC-OCO uses the $\tau$-slot delayed gradient $\nabla{f}_{t-\tau}(\mathbf{x}_{t-\tau})$ in \textbf{P2} for accumulated loss minimization. Compared with the original $\textbf{P1}$, the long-term constraints (\ref{eq:ltc}) are converted to penalizing $\mathbf{g}_{t-\tau}(\mathbf{x})$ for queue stabilities as one part of the objective in $\textbf{P2}$. Note that DTC-OCO uses a novel constraint penalty with double regularization $\alpha\Vert\mathbf{x}-\mathbf{x}_{t-\tau}\Vert_2^2$ and $\eta\Vert\mathbf{x}-\mathbf{x}_{t-1}\Vert_2^2$ in the per-slot objective in $\textbf{P2}$ to handle the asynchrony between information feedback and decision updates. This will be shown, analytically in Sections \ref{Subsec:Regret Bound} and \ref{Subsec:Discussions on the Performance Bounds} and numerically in Section \ref{Sec:Application to Network Resource Allocation}, to give DTC-OCO substantial performance advantage over existing works in terms of regret bounds and average performance.
   
To summarize, DTC-OCO is comprised of three major steps: 1) Initialize $\mathbf{x}_t\in\mathcal{X}_0,\mathbf{Q}_t=\mathbf{0},\forall{t}\in[1,\tau]$
and $\mathbf{g}_0(\mathbf{x})\equiv\mathbf{0}$; 2) At the beginning of each slot $t>\tau$, obtain the current decision $\mathbf{x}_t$ by solving  $\textbf{P2}$; 3) At the end of each slot $t>\tau$, update the virtual queue $\mathbf{Q}_t$ via (\ref{eq:VQ}).\footnote{If information feedback of $f_t(\mathbf{x})$ and $\mathbf{g}_t(\mathbf{x})$ are respectively delayed for $\tau_1$ and $\tau_2$ slots with $\tau_1\neq\tau_2$, DTC-OCO can still be applied by setting $\tau=\max\{\tau_1,\tau_2\}$.} Note that DTC-OCO has three step-size parameters $\alpha,\eta,\gamma>0$. Their choice depends on our knowledge of the system and will be discussed in Section~\ref{Subsec:Discussions on the Performance Bounds}, after we derive the regret and constraint violation bounds in Section~\ref{Subsec:Regret Bound}. We will clarify the impact of these step-sizes on those bounds.

The main difference between DTC-OCO and the saddle-point-typed OCO algorithms \cite{Trade,LTC-Toff,T.Chen,X.CaoTau} is that DTC-OCO uses a virtual queue to track the constraint violation. The virtual queue was also used in Lyapunov optimization~\cite{Neely}, and later extended to OCO in \cite{LTC-HY}, \cite{Yu-SC}, and \cite{X.Cao}. Although a small part of our performance bounding borrows some techniques from Lyapunov drift analysis, as explained in Section II, DTC-OCO is structurally different from Lyapunov optimization. Furthermore, the virtual-queue-based OCO algorithms in \cite{LTC-HY} and \cite{Yu-SC} were designed for time-invariant and i.i.d. constraints, respectively, while obtaining only static regret bounds. Compared with \cite{X.Cao}, DTC-OCO only uses the gradient of the loss functions \textit{at the past decision points}, instead of the complete information of the past loss functions. Furthermore, \cite{LTC-HY},~\cite{Yu-SC},~and~\cite{X.Cao} are limited to one-slot feedback delay. Therefore, the virtual queue construction, algorithm design, and performance bound derivation for DTC-OCO are all substantially different from those of \cite{LTC-HY},~\cite{Yu-SC},~and~\cite{X.Cao}.

\subsection{Regret and Constraint Violation Bounds}
\label{Subsec:Regret Bound}

In this section, we present new techniques to derive the performance bounds of DTC-OCO, particularly to account for its constraint penalty with double regularization.

We make the following  assumptions that are common in the literature for constrained OCO \cite{Trade}\nocite{LTC-Toff}\nocite{LTC-HY}\nocite{Yu-SC}\nocite{T.Chen}-\cite{X.Cao}, \cite{X.CaoTau}.

\begin{assumption}
The gradient $\nabla{f}_t(\mathbf{x})$ is bounded: $\exists{D}\!>\!0$, \st
\begin{align}
        \Vert\nabla f_t(\mathbf{x})\Vert_2\le D,\quad\forall{\mathbf{x}}\in\mathcal{X}_0,\quad\forall{t}.\label{eq:D}
\end{align}
\end{assumption}

\begin{assumption} For any $t$, $\mathbf{g}_t(\mathbf{x})$ satisfies
the following:

2.1) $\mathbf{g}_t(\mathbf{x})$ is Lipschitz continuous on $\mathcal{X}_0$: $\exists\beta>0$, \st
\begin{align}
        \Vert\mathbf{g}_t(\mathbf{x})-\mathbf{g}_t(\mathbf{y})\Vert_2\le\beta\Vert\mathbf{x}-\mathbf{y}\Vert_2,\quad\forall\mathbf{x},\mathbf{y}\in\mathcal{X}_0,\quad\forall{t}.\label{eq:Beta}
\end{align}

2.2) $\mathbf{g}_t(\mathbf{x})$ is bounded: $\exists{G}>0$, \st 
\begin{align}
        \Vert \mathbf{g}_t(\mathbf{x})\Vert_2\leq G,\quad\forall{\mathbf{x}}\in\mathcal{X}_0,\quad\forall{t}.\label{eq:G}
\end{align}

2.3) Existence of an interior point: $\exists\epsilon\!>\!0$ and $\tilde{\mathbf{x}}_t\in\mathcal{X}_0$, \st
\begin{align}
        \mathbf{g}_{t}(\tilde{\mathbf{x}}_t)\preceq-\epsilon\mathbf{1},\quad\forall{t}.\label{eq:epsilon}
\end{align}
\end{assumption}

\begin{assumption}
The radius of $\mathcal{X}_0$ is bounded: $\exists{R}>0$, \st 
\begin{align}
        \Vert\mathbf{x}-\mathbf{y}\Vert_2\leq R,\quad\forall\mathbf{x},\mathbf{y}\in\mathcal{X}_0.\label{eq:R}
\end{align}
\end{assumption}

We first provide bounds on the virtual queue vector in the following lemma. The proof follows from the virtual queue definition in (\ref{eq:VQ}) and is omitted due to page limitation.
\begin{lemma}\label{lm:VQ}
The virtual queue vector generated by DTC-OCO is bounded for any $t>\tau$ by the following inequalities:
\begin{align}
        &\mathbf{Q}_t+\gamma\mathbf{g}_{t-\tau}(\mathbf{x}_{t})\succeq\mathbf{0},\label{eq:VQ2}\\
        &\Vert{\mathbf{Q}}_t\Vert_2\geq\Vert\gamma\mathbf{g}_{t-\tau}(\mathbf{x}_{t})\Vert_2,\label{eq:VQ3}\\
        &\Vert\mathbf{Q}_t\Vert_2\leq \Vert\mathbf{Q}_{t-1}\Vert_2+\Vert\gamma\mathbf{g}_{t-\tau}(\mathbf{x}_{t})\Vert_2.\label{eq:VQ4}
\end{align}
\end{lemma}

Define $L_t\triangleq\frac{1}{2}\Vert\mathbf{Q}_t\Vert_2^2$ as a quadratic Lyapunov function and $\Delta_t\triangleq L_{t+1}-L_{t}$ as the corresponding Lyapunov drift \cite{Neely}. Leveraging results in Lemma~\ref{lm:VQ}, we provide an upper bound on $\Delta_t$ in the following lemma.

\begin{lemma}\label{lm:drift}
The Lyapunov drift is upper bounded for any $t>\tau$ as follows:
\begin{align}
        \Delta_{t-1}&\leq\gamma\mathbf{Q}_{t-1}^T\mathbf{g}_{t-\tau}(\mathbf{x}_{t})+\Vert\gamma\mathbf{g}_{t-\tau}(\mathbf{x}_{t})\Vert_2^2.\label{eq:drift}
\end{align}
\end{lemma}

\textit{Proof:} (\textit{Proof outline due to page limitation}) For any $c\in\mathcal{C}$ and $t>\tau$, we can verify that
\begin{align*}
        \frac{1}{2}(Q_t^c)^2-\frac{1}{2}(Q_{t-1}^c)^{2}\le\gamma{Q}_{t-1}^cg_{t-\tau}^c(\mathbf{x}_{t})+[\gamma{g}_{t-\tau}^c(\mathbf{x}_{t})]^2
\end{align*}
by considering the two cases from (\ref{eq:VQ}): 1) $Q_{t-1}^c+\gamma{g}_{t-\tau}^c(\mathbf{x}_{t})\ge-\gamma{g}_{t-\tau}^c(\mathbf{x}_{t})$ and 2) $-\gamma{g}_{t-\tau}^c(\mathbf{x}_{t})>Q_{t-1}^c+\gamma{g}_{t-\tau}^c(\mathbf{x}_{t})$. Summing the above inequalities over $c\in\mathcal{C}$ yields (\ref{eq:drift}).
\endIEEEproof

We also require the following lemma, which is reproduced from Lemma 2.8 in \cite{OLBK}.
\begin{lemma}\label{lm:StrConv}
Let $\mathcal{S}\in\mathbb{R}^n$ be a nonempty convex set. Let $h(\mathbf{v}):\mathbb{R}^n\to\mathbb{R}$ be a $\varrho$-strongly-convex function over $\mathcal{S}$ w.r.t. a norm $\Vert\cdot\Vert$. Let $\mathbf{w}=\arg\min_{\mathbf{v}\in\mathcal{S}}h(\mathbf{v})$. Then, for any $\mathbf{u}\in\mathcal{S}$, we have $h(\mathbf{w})\le{h}(\mathbf{u})-\frac{\varrho}{2}\Vert\mathbf{u}-\mathbf{w}\Vert^2$.
\end{lemma}

A main goal of this paper is to examine the impact of multi-slot feedback delay on the dynamic regret  bound for constrained OCO, which has not been addressed in the existing literature. To this end, we need to quantify the accumulated variations of the underlying time-varying system. We define the accumulated variation of the dynamic benchmark $\{\mathbf{x}_t^\star\}$ (commonly termed the path length \cite{Zinkevich}) as
\begin{align}
        \Delta_{\mathbf{x}^\star}\triangleq\sum_{t=1}^{T}\Vert\mathbf{x}_{t}^{\star}-\mathbf{x}_{t-1}^{\star}\Vert_2.\label{eq:tmpx}
\end{align}
Furthermore, we define the accumulated squared variation of the constraint function sequence $\{\mathbf{g}_t(\mathbf{x})\}$ as
\begin{align}
        \Delta_{\mathbf{g}}\triangleq\sum_{t=1}^{T}\max_{\mathbf{x}\in\mathcal{X}_0}\left\{\Vert\mathbf{g}_t(\mathbf{x})-\mathbf{g}_{t-1}(\mathbf{x})\Vert_2^2\right\}.\label{eq:tmpg}
\end{align}
Another related quantity regarding the accumulated variation of $\{\mathbf{g}_t(\mathbf{x})\}$ is defined as
\begin{align}
        \tilde{\Delta}_{\mathbf{g}}\triangleq\sum_{t=1}^{T}\max_{\mathbf{x}\in\mathcal{X}_0}\left\{\left\Vert\mathbf{g}_{t}(\mathbf{x})-\mathbf{g}_{t-1}(\mathbf{x})\right\Vert_2\right\}.\label{tmp-g2}
\end{align}
In the order sense, $\Delta_{\mathbf{g}}$ is usually smaller than $\tilde{\Delta}_{\mathbf{g}}$
for a constraint function sequence $\{\mathbf{g}_t(\mathbf{x})\}$ that varies sublinearly~\cite{X.Cao}.

Leveraging results in Lemmas~\ref{lm:VQ} and \ref{lm:drift}, and the tuning freedom brought by the double regularization for constructing telescoping terms, the following theorem provides an upper bound on the dynamic regret $\text{RE}_{\text{d}}(T)$ for DTC-OCO with $\tau$-slot feedback delay. 

\setcounter{theorem}{3}
\begin{theorem}\label{thm:reg}
For any $\alpha,\gamma>0$ and $\eta\ge\gamma^2\beta^2$, the dynamic regret of DTC-OCO is upper bounded by
\begin{align}
        \text{RE}_{\text{d}}(T)&\le\frac{D^2}{4\alpha}T+\frac{\gamma^2G^2}{2}+\gamma^2\Delta_{\mathbf{g}}\notag\\
        &\quad+(\alpha\tau+\eta)(R^2+2R\Delta_{\mathbf{x}^\star})+DR\tau.\label{eq:reg}
\end{align}
\end{theorem}

\textit{Proof:} The objective function of $\textbf{P2}$ is $2(\alpha+\eta)$-strongly-convex over $\mathcal{X}_0$ w.r.t. Euclidean norm $\Vert\cdot\Vert_2$ due to the double regularization. Since $\mathbf{x}_t$ minimizes $\textbf{P2}$ over $\mathcal{X}_0$ for any $t>\tau$, we have
\begin{align}
        &[\nabla f_{t-\tau}(\mathbf{x}_{t-\tau})]^T(\mathbf{x}_t-\mathbf{x}_{t-\tau})+\alpha\Vert\mathbf{x}_t-\mathbf{x}_{t-\tau}\Vert_2^2\notag\\
        &\quad+[\mathbf{Q}_{t-1}+\gamma\mathbf{g}_{t-\tau-1}(\mathbf{x}_{t-1})]^T[\gamma\mathbf{g}_{t-\tau}(\mathbf{x}_{t})]\!+\!\eta\Vert\mathbf{x}_t-\mathbf{x}_{t-1}\Vert_2^2\notag\\
        &\stackrel{(a)}{\le} [\nabla f_{t-\tau}(\mathbf{x}_{t-\tau})]^T(\mathbf{x}_{t-\tau}^\star-\mathbf{x}_{t-\tau})+\alpha\Vert\mathbf{x}_{t-\tau}^\star-\mathbf{x}_{t-\tau}\Vert_2^2\notag\\
        &\quad+[\mathbf{Q}_{t-1}+\gamma\mathbf{g}_{t-\tau-1}(\mathbf{x}_{t-1})]^T[\gamma\mathbf{g}_{t-\tau}(\mathbf{x}_{t-\tau}^\star)]\notag\\
        &\quad+\eta\Vert\mathbf{x}_{t-\tau}^\star-\mathbf{x}_{t-1}\Vert_2^2-(\alpha+\eta)\Vert\mathbf{x}_t-\mathbf{x}_{t-\tau}^\star\Vert_2^2\label{eq:thm1-0}\\
        &\stackrel{(b)}{\le} [\nabla f_{t-\tau}(\mathbf{x}_{t-\tau})]^T(\mathbf{x}_{t-\tau}^\star-\mathbf{x}_{t-\tau})\notag\\
        &\quad+\alpha(\Vert\mathbf{x}_{t-\tau}^\star-\mathbf{x}_{t-\tau}\Vert_2^2-\Vert\mathbf{x}_t-\mathbf{x}_{t-\tau}^\star\Vert_2^2)\notag\\
        &\quad+\eta(\Vert\mathbf{x}_{t-\tau}^\star-\mathbf{x}_{t-1}\Vert_2^2-\Vert\mathbf{x}_t-\mathbf{x}_{t-\tau}^\star\Vert_2^{2}),\label{eq:thm1-1}
\end{align}
where $(a)$ follows from Lemma~\ref{lm:StrConv}; and $(b)$ is because $\mathbf{Q}_\tau=\mathbf{0}$, $\mathbf{g}_0(\mathbf{x})\equiv\mathbf{0}$ by initialization, $\mathbf{Q}_t+\gamma\mathbf{g}_{t-\tau}(\mathbf{x}_{t})\succeq\mathbf{0},\forall{t}>\tau$, in (\ref{eq:VQ2}), $\gamma>0$, and $\mathbf{g}_{t-\tau}(\mathbf{x}_{t-\tau}^\star)\preceq\mathbf{0},\forall{t}>\tau$,
in (\ref{eq:xtstar}), such that $[\mathbf{Q}_{t-1}+\gamma\mathbf{g}_{t-\tau-1}(\mathbf{x}_{t-1})]^T[\gamma\mathbf{g}_{t-\tau}(\mathbf{x}_{t-\tau}^\star)]\le0,\forall{t}>\tau$.

Now, we bound the second and third terms in (\ref{eq:thm1-1}). From $\Vert\mathbf{a}+\mathbf{b}\Vert_2^2\geq\Vert\mathbf{a}\Vert_2^2+\Vert\mathbf{b}\Vert_2^2-2\Vert\mathbf{a}\Vert_2\Vert\mathbf{b}\Vert_2$, we have
\begin{align}
        &\Vert\mathbf{x}_{t-\tau}^\star-\mathbf{x}_{t-\tau}\Vert_2^2-\Vert\mathbf{x}_t-\mathbf{x}_{t-\tau}^\star\Vert_2^2\notag\\
        &\le\Vert\mathbf{x}_{t-\tau}^\star-\mathbf{x}_{t-\tau}\Vert_2^2-\Vert\mathbf{x}_{t}^\star-\mathbf{x}_t\Vert_2^2-\Vert\mathbf{x}_{t-\tau}^\star-\mathbf{x}_{t}^\star\Vert_2^2\notag\\
        &\quad+2\Vert\mathbf{x}_{t}^\star-\mathbf{x}_t\Vert_2\Vert\mathbf{x}_{t-\tau}^\star-\mathbf{x}_{t}^\star\Vert_2\le\Phi_{t-\tau}+2R\phi_{t-\tau},\label{eq:thm1-2}
\end{align}
where $\Phi_{t-\tau}\triangleq\Vert\mathbf{x}_{t-\tau}^\star-\mathbf{x}_{t-\tau}\Vert_2^2-\Vert\mathbf{x}_{t}^\star-\mathbf{x}_t\Vert_2^2$ and $\phi_{t-\tau}\triangleq\Vert\mathbf{x}_{t-\tau}^\star-\mathbf{x}_{t}^\star\Vert_2$. Similarly, we can show that
\begin{align}
        \Vert\mathbf{x}_{t-\tau}^\star-\mathbf{x}_{t-1}\Vert_2^2-\Vert\mathbf{x}_t-\mathbf{x}_{t-\tau}^\star\Vert_2^2
        \le\Psi_{t-\tau}+2R\psi_{t-\tau},\label{eq:thm1-2'}
\end{align}
where $\Psi_{t-\tau}\triangleq\Vert\mathbf{x}_{t-\tau}^\star-\mathbf{x}_{t-1}\Vert_2^2-\Vert\mathbf{x}_{t-\tau+1}^\star-\mathbf{x}_t\Vert_2^2$
and $\psi_{t-\tau}\triangleq\Vert\mathbf{x}_{t-\tau}^\star-\mathbf{x}_{t-\tau+1}^\star\Vert_2$.

Substituting (\ref{eq:thm1-2}) and (\ref{eq:thm1-2'}) into (\ref{eq:thm1-1}),
adding $f_{t-\tau}(\mathbf{x}_{t-\tau})$ on both sides, applying the first-order condition of convexity
\begin{align*}
        f_{t-\tau}(\mathbf{x}_{t-\tau})+[\nabla f_{t-\tau}(\mathbf{x}_{t-\tau})]^T(\mathbf{x}_{t-\tau}^\star-\mathbf{x}_{t-\tau})\le{f}_{t-\tau}(\mathbf{x}_{t-\tau}^\star)
\end{align*}
to its RHS, and rearranging terms, we have
\begin{align}
        &f_{t-\tau}(\mathbf{x}_{t-\tau})-f_{t-\tau}(\mathbf{x}_{t-\tau}^\star)\notag\\
        &\le-[\nabla f_{t-\tau}(\mathbf{x}_{t-\tau})]^T(\mathbf{x}_t-\mathbf{x}_{t-\tau})-\alpha\Vert\mathbf{x}_t-\mathbf{x}_{t-\tau}\Vert_2^2\notag\\
        &\quad-[\mathbf{Q}_{t-1}+\gamma\mathbf{g}_{t-\tau-1}(\mathbf{x}_{t-1})]^T[\gamma\mathbf{g}_{t-\tau}(\mathbf{x}_{t})]\!-\!\eta\Vert\mathbf{x}_t-\mathbf{x}_{t-1}\Vert_2^2\notag\\
        &\quad+\alpha(\Phi_{t-\tau}+2R\phi_{t-\tau})+\eta(\Psi_{t-\tau}+2R\psi_{t-\tau}).\label{eq:thm1-4}
\end{align}

We now bound the right-hand side of (\ref{eq:thm1-4}). Note that
\begin{align}
        &-[\mathbf{Q}_{t-1}+\gamma\mathbf{g}_{t-\tau-1}(\mathbf{x}_{t-1})]^T[\gamma\mathbf{g}_{t-\tau}(\mathbf{x}_{t})]\notag\\
        &\stackrel{(a)}{\le}-\Delta_{t-1}+\Vert\gamma\mathbf{g}_{t-\tau}(\mathbf{x}_t)\Vert_2^2-\gamma^2\mathbf{g}_{t-\tau-1}^T(\mathbf{x}_{t-1})\mathbf{g}_{t-\tau}(\mathbf{x}_{t})\notag\\
        &\stackrel{(b)}{=}-\Delta_{t-1}+\frac{\gamma^2}{2}(\Vert\mathbf{g}_{t-\tau}(\mathbf{x}_t)\Vert_2^2-\Vert\mathbf{g}_{t-\tau-1}(\mathbf{x}_{t-1})\Vert_2^2)\notag\\
        &\quad+\frac{\gamma^2}{2}\Vert\mathbf{g}_{t-\tau}(\mathbf{x}_t)-\mathbf{g}_{t-\tau-1}(\mathbf{x}_{t-1})\Vert_2^2\notag\\
        &\stackrel{(c)}{\le}-\Delta_{t-1}+\gamma^2\left(\frac{1}{2}\varphi_{t-\tau}+\beta^2\Vert\mathbf{x}_t-\mathbf{x}_{t-1}\Vert_2^2+\varpi_{t-\tau}\right)\!,\!\!\!\label{eq:thm1-5}
\end{align}
where $\varphi_{t-\tau}\triangleq \Vert\mathbf{g}_{t-\tau}(\mathbf{x}_t)\Vert_2^2-\Vert\mathbf{g}_{t-\tau-1}(\mathbf{x}_{t-1})\Vert_2^2$
and $\varpi_{t-\tau}\triangleq\Vert\mathbf{g}_{t-\tau}(\mathbf{x}_{t-1})-\mathbf{g}_{t-\tau-1}(\mathbf{x}_{t-1})\Vert_2^2$. Here, $(a)$ follows from rearranging terms of (\ref{eq:drift}) in Lemma~\ref{lm:drift} such that $-\gamma\mathbf{Q}_{t-1}^T\mathbf{g}_{t-\tau}(\mathbf{x}_t)\le-\Delta_{t-1}+\Vert\gamma\mathbf{g}_{t-\tau}(\mathbf{x}_{t})\Vert_2^2$,
$(b)$ is because $\mathbf{a}^T\mathbf{b}=\frac{1}{2}(\Vert\mathbf{a}\Vert_2^2+\Vert\mathbf{b}\Vert_2^2-\Vert\mathbf{a}-\mathbf{b}\Vert_2^2)$,
and $(c)$ follows from $\mathbf{g}_t(\mathbf{x})$ being Lipschitz continuous in (\ref{eq:Beta}) and the fact that $\frac{1}{2}\Vert\mathbf{a}+\mathbf{b}\Vert_2^2\leq\Vert\mathbf{a}\Vert_2^2+\Vert\mathbf{b}\Vert_2^2$.

Substituting (\ref{eq:thm1-5}) into (\ref{eq:thm1-4}), we have
\begin{align}
        &f_{t-\tau}(\mathbf{x}_{t-\tau})-f_{t-\tau}(\mathbf{x}_{t-\tau}^\star)\notag\\
        &\le-[\nabla f_{t-\tau}(\mathbf{x}_{t-\tau})]^T(\mathbf{x}_t-\mathbf{x}_{t-\tau})-\alpha\Vert\mathbf{x}_t-\mathbf{x}_{t-\tau}\Vert_2^2\notag\\
        &\quad+(\gamma^2\beta^2-\eta)\Vert\mathbf{x}_t-\mathbf{x}_{t-1}\Vert_2^2-\Delta_{t-1}+\frac{\gamma^2}{2}\varphi_{t-\tau}+\gamma^2\varpi_{t-\tau}\notag\\
        &\quad+\alpha(\Phi_{t-\tau}+2R\phi_{t-\tau})+\eta(\Psi_{t-\tau}+2R\psi_{t-\tau})\notag\\
        &\stackrel{(a)}{\le} \frac{D^2}{4\alpha}-\Delta_{t-1}+\frac{\gamma^2}{2}\varphi_{t-\tau}+\gamma^2\varpi_{t-\tau}\notag\\
        &\quad+\alpha(\Phi_{t-\tau}+2R\phi_{t-\tau})+\eta(\Psi_{t-\tau}+2R\psi_{t-\tau}),\label{eq:thm1-6}
\end{align}
where $(a)$ follows from $\eta\ge\gamma^2\beta^2$, the bound on $\nabla{f}_t(\mathbf{x})$ in (\ref{eq:D}), and completing
the square such that
\begin{align*}
        &-[\nabla f_{t-\tau}(\mathbf{x}_{t-\tau})]^T(\mathbf{x}_t-\mathbf{x}_{t-\tau})-\alpha\Vert\mathbf{x}_t-\mathbf{x}_{t-\tau}\Vert_2^2\\
        &=\!-\!\left\Vert\frac{\nabla f_{t-\tau}(\mathbf{x}_{t-\tau})}{2\sqrt{\alpha}}\!+\!\sqrt{\alpha}(\mathbf{x}_t\!-\mathbf{\!x}_{t-\tau})\right\Vert_2^2\!\!\!+\!\frac{1}{4\alpha}\Vert\nabla{f}_{t-\tau}(\mathbf{x}_{t-\tau})\Vert_2^2\notag\\
        &\leq\frac{1}{4\alpha}\Vert\nabla f_{t-\tau}(\mathbf{x}_{t-\tau})\Vert_2^2\le\frac{D^2}{4\alpha}.
\end{align*}

Summing $(\ref{eq:thm1-6})$ over $t\in[\tau+1,T]$, we have
\begin{align}
        &\sum_{t=\tau+1}^T f_{t-\tau}(\mathbf{x}_{t-\tau})-f_{t-\tau}(\mathbf{x}_{t-\tau}^\star)=\sum_{t=1}^{T-\tau}f_t(\mathbf{x}_t)-f_{t}(\mathbf{x}_t^\star)\notag\\
        &\stackrel{(a)}{\le}\frac{D^2}{4\alpha}T+\frac{\gamma^2G^2}{2}+\gamma^2\Delta_{\mathbf{g}}+(\alpha\tau+\eta)(R^2+2R\Delta_{\mathbf{x}^\star}),\!\label{eq:thm1-7}
\end{align}
where $(a)$ follows from $\Delta_{t-1}$, $\varphi_{t-\tau}$, $\varpi_{t-\tau}$,
$\Phi_{t-\tau}$, $\phi_{t-\tau}$, $\Psi_{t-\tau}$ and $\psi_{t-\tau}$ all
being telescoping terms such that their sums over $t\in[\tau+1,T]$ are upper bounded by $0$, $G^2$, $\Delta_{\mathbf{g}}$, $\tau{R}^2$, $\tau\Delta_{\mathbf{x}^\star}$, $R^2$, and $\Delta_{\mathbf{x}^\star}$, respectively.

Finally, adding $\sum_{t=T-\tau+1}^T f_t(\mathbf{x}_t)-f_{t}(\mathbf{x}_t^\star)$ on both sides of (\ref{eq:thm1-7}), and noting that the convexity of $f_t(\mathbf{x})$ implies
\begin{align}
        f_t(\mathbf{x}_t)-f_{t}(\mathbf{x}_t^\star)\le\Vert\nabla f_t(\mathbf{x}_t)\Vert_2\Vert\mathbf{x}_t^\star-\mathbf{x}_t\Vert_2\le{D}R,\label{eq:thm1-8}
\end{align} 
we complete the proof. 
\endIEEEproof

Next, leveraging the proof techniques for the dynamic regret $\text{RE}_{\text{d}}(T)$ in Theorem~\ref{thm:reg}, we provide an upper bound on the static regret $\text{RE}_{\text{s}}(T)$ yielded by DTC-OCO.

\begin{theorem}\label{thm:staticreg}
For any $\alpha,\gamma>0$ and $\eta\ge\gamma^2\beta^2$, the static
regret of DTC-OCO is upper bounded by
\begin{align}
        \text{RE}_{\text{s}}(T)\le\frac{D^2}{4\alpha}T\!+\!\frac{\gamma^2G^2}{2}\!+\!\gamma^2\Delta_{\mathbf{g}}\!+\!(\alpha\tau\!+\!\eta)R^2\!+\!DR\tau.
\end{align}
\end{theorem}
\textit{Proof}: (\textit{Proof outline due to page limitation}) Replacing all the per-slot optimizers with the offline benchmark $\mathbf{x}^\star$ in the proof of Theorem \ref{thm:reg}, we can show that (\ref{eq:thm1-6}) still holds by redefining $\Phi_{t-\tau}\triangleq\Vert\mathbf{x}^\star-\mathbf{x}_{t-\tau}\Vert_2^2-\Vert\mathbf{x}^\star-\mathbf{x}_t\Vert_2^2$, $\phi_{t-\tau}\triangleq0$, $\Psi_{t-\tau}\triangleq\Vert\mathbf{x}^\star-\mathbf{x}_{t-1}\Vert_2^2-\Vert\mathbf{x}^\star-\mathbf{x}_t\Vert_2^2$, and $\psi_{t-\tau}\triangleq0$. Summing the above version of (\ref{eq:thm1-6}) over $t\in[\tau+1,T]$, noting that $\Phi_{t-\tau}$ and $\Psi_{t-\tau}$ are still telescoping, and leveraging (\ref{eq:thm1-8}), we complete the proof.\endIEEEproof

We now proceed to provide an upper bound on the constraint violation $\text{VO}^c(T)$ for DTC-OCO. We first relate the virtual queue vector $\mathbf{Q}_T$ to $\text{VO}^c(T)$ in the following lemma.

\setcounter{lemma}{5}
\begin{lemma}\label{lm:VQVio}
The virtual queue vector produced by DTC-OCO satisfies the following inequality for any $c\in\mathcal{C}$:
\begin{align}
        \text{VO}^c(T)\le\frac{1}{\gamma}\Vert\mathbf{Q}_T\Vert_2+\tau\tilde{\Delta}_{\mathbf{g}}+G\tau.\label{eq:VioQT}
\end{align}
\end{lemma}
\textit{Proof:} From (\ref{eq:VQ}), we have $Q_{t}^c\ge Q_{t-1}^{c}+\gamma{g}_{t-\tau}^c(\mathbf{x}_{t}),\forall{t}>\tau$. Summing it over $t\in[\tau+1,T]$ and rearranging terms, we have
\begin{align*}
        \sum_{t=\tau+1}^Tg_{t-\tau}^c(\mathbf{x}_t)\le\frac{1}{\gamma}\sum_{t=\tau+1}^TQ_t^c-Q_{t-1}^c=\frac{1}{\gamma}Q_T^c.%\label{eq:thm2-2}
\end{align*}
From the above inequality and the definition of $\text{VO}^c(T)$ in (\ref{eq:Vio}), we have
\begin{align*}
        \text{VO}^c(T)\le\frac{1}{\gamma}Q_T^c+\!\sum_{t=1}^{T-\tau}[g_{t+\tau}^c(\mathbf{x}_{t+\tau})\!-g_{t}^c(\mathbf{x}_{t+\tau})]+\!\sum_{t=1}^{\tau}g_t^c(\mathbf{x}_t).
\end{align*}
Noting  $\Vert\mathbf{a}\Vert_\infty\le\Vert\mathbf{a}\Vert_2$, the bound on $\mathbf{g}_t(\mathbf{x})$ in (\ref{eq:G}), and the definition of $\tilde{\Delta}_{\mathbf{g}}$ in (\ref{tmp-g2}), we complete the proof.
\endIEEEproof

From Lemma \ref{lm:VQVio}, we see that one can bound the constraint violation $\text{VO}^c(T)$ through an upper bound on the virtual queue vector $\mathbf{Q}_T$. This is used to obtain an upper bound on the constraint violation for DTC-OCO in the following theorem.

\setcounter{theorem}{6}
\begin{theorem}\label{thm:vio}
For any $\alpha,\eta,\gamma>0$, the constraint
violation of DTC-OCO for any $c\in\mathcal{C}$ is upper bounded by
\begin{align}
        \text{VO}^c(T)\le\!2G\!+\!\frac{2\gamma^2G^2\!+\!DR\!+\!(\alpha\!+\!\eta)R^2}{\epsilon\gamma^2}\!+\!\tau\tilde{\Delta}_{\mathbf{g}}\!+\!G\tau.\label{eq:vio}
\end{align}
\end{theorem}
\textit{Proof}: From Lemma~\ref{lm:StrConv}, we can show that inequality (\ref{eq:thm1-0}) still holds for any $t>\tau$ after replacing the per-slot optimizer $\mathbf{x}_{t-\tau}^\star$ with the interior point $\tilde{\mathbf{x}}_{t-\tau}$. We have
\begin{align}
        &[\mathbf{Q}_{t-1}+\gamma\mathbf{g}_{t-\tau-1}(\mathbf{x}_{t-1})]^T[\gamma\mathbf{g}_{t-\tau}(\tilde{\mathbf{x}}_{t-\tau})]\notag\\
        &\stackrel{(a)}{\le}-\epsilon\gamma[\mathbf{Q}_{t-1}+\gamma\mathbf{g}_{t-\tau-1}(\mathbf{x}_{t-1})]^T\mathbf{1}\notag\\
        &\stackrel{(b)}{\le}-\epsilon\gamma\Vert\mathbf{Q}_{t-1}+\gamma\mathbf{g}_{t-\tau-1}(\mathbf{x}_{t-1})\Vert_2\notag\\
        &\stackrel{(c)}{\le}-\epsilon\gamma(\Vert\mathbf{Q}_{t-1}\Vert_2-\Vert\gamma\mathbf{g}_{t-\tau-1}(\mathbf{x}_{t-1})\Vert_2),\label{eq:thm2-6}
\end{align}
where $(a)$ follows from the existence of interior point in (\ref{eq:epsilon}) and the virtual queue bound in (\ref{eq:VQ2}), $(b)$ is because $\Vert\mathbf{a}\Vert_2\le\Vert\mathbf{a}\Vert_1$, and $(c)$ follows from $|\Vert\mathbf{a}\Vert_2-\Vert\mathbf{b}\Vert_2|\le\Vert\mathbf{a}-\mathbf{b}\Vert_2$.
Applying (\ref{eq:thm2-6}) to the aforementioned version of (\ref{eq:thm1-0}) with $\tilde{\mathbf{x}}_{t-\tau}$ and rearranging terms, we have
\begin{align}
        &\gamma\mathbf{Q}_{t-1}^T\mathbf{g}_{t-\tau}(\mathbf{x}_t)\notag\\
        &\le-\epsilon\gamma(\Vert\mathbf{Q}_{t-1}\Vert_2-\Vert\gamma\mathbf{g}_{t-\tau-1}(\mathbf{x}_{t-1})\Vert_2)-\alpha\Vert\mathbf{x}_t-\mathbf{x}_{t-\tau}\Vert_2^2\notag\\
        &\quad-[\gamma\mathbf{g}_{t-\tau-1}(\mathbf{x}_{t-1})]^T[\gamma\mathbf{g}_{t-\tau}(\mathbf{x}_t)]-\eta\Vert\mathbf{x}_t-\mathbf{x}_{t-1}\Vert_2^2\notag\\
        &\quad+[\nabla{f}_{t-\tau}(\mathbf{x}_{t-\tau})]^T(\tilde{\mathbf{x}}_{t-\tau}-\mathbf{x}_{t})+\alpha\Vert\tilde{\mathbf{x}}_{t-\tau}-\mathbf{x}_{t-\tau}\Vert_2^2\notag\\
        &\quad+\eta\Vert\tilde{\mathbf{x}}_{t-\tau}-\mathbf{x}_{t-1}\Vert_2^2-(\alpha+\eta)\Vert\mathbf{x}_t-\tilde{\mathbf{x}}_{t-\tau}\Vert_2^2\notag\\
        &\stackrel{(a)}{\le}-\epsilon\gamma\Vert\mathbf{Q}_{t-1}\Vert_2+\epsilon\gamma^2G+\gamma^2G^2+DR+(\alpha+\eta)R^2,\!\!\label{eq:thm2-7}
\end{align}
where $(a)$ follows from the Cauchy-Schwartz inequality $|\mathbf{a}^T\mathbf{b}|\le\Vert\mathbf{a}\Vert_2\Vert\mathbf{b}\Vert_2$, the bound on $\nabla{f}_t(\mathbf{x})$ in (\ref{eq:D}), the bound on $\mathbf{g}_t(\mathbf{x})$ in  (\ref{eq:G}), and the bound on $\mathcal{X}_0$ in (\ref{eq:R}). Substituting (\ref{eq:thm2-7})
into (\ref{eq:drift}) in Lemma \ref{lm:drift} and noting that $\Vert\mathbf{g}_{t-\tau}(\mathbf{x}_{t})\Vert_2^2\le{G}^2$
from (\ref{eq:G}) yields
\begin{align*}
        \Delta_{t-1}&\leq-\epsilon\gamma\Vert\mathbf{Q}_{t-1}\Vert_2+\epsilon\gamma^2G+2\gamma^2G^2+DR+(\alpha+\eta)R^2.
\end{align*}

Thus, a sufficient condition for $\Delta_{t-1}<0$ is 
\begin{align}
        \Vert\mathbf{Q}_{t-1}\Vert_2>\gamma{G}+\frac{2\gamma^2G^2+DR+(\alpha+\eta)R^2}{\epsilon\gamma}.\label{eq:thm2-8}
\end{align}
If (\ref{eq:thm2-8}) holds, we have $\Vert\mathbf{Q}_{t}\Vert_2<\Vert\mathbf{Q}_{t-1}\Vert_2$,
\ie the virtual queue decreases; otherwise, from the virtual queue bound in (\ref{eq:VQ4}), there is
a maximum increase from $\Vert\mathbf{Q}_{t-1}\Vert_2$ to $\Vert\mathbf{Q}_t\Vert_2$ since $\Vert\mathbf{Q}_t\Vert_2-\Vert\mathbf{Q}_{t-1}\Vert_2\leq \Vert\gamma\mathbf{g}_{t-\tau}(\mathbf{x}_{t})\Vert_2\le\gamma{G}$.
Therefore, the virtual queue is upper bounded for any $t>\tau$ by
\begin{align}
        \Vert\mathbf{Q}_t\Vert_2\le2\gamma{G}+\frac{2\gamma^2G^2+DR+(\alpha+\eta)R^2}{\epsilon\gamma}\label{eq:thm2-9}.
\end{align}
Substituting (\ref{eq:thm2-9}) into (\ref{eq:VioQT}), we complete the proof. \endIEEEproof

\subsection{Discussion on the Regret and Constraint Violation Bounds}
\label{Subsec:Discussions on the Performance Bounds}

In this section, we discuss the sufficient conditions for DTC-OCO to yield sublinear regret and constraint violation, and highlight several prominent advantages of DTC-OCO over existing constrained OCO algorithms.

\subsubsection{Sublinear Regret and Constraint Violation}

From Theorems~\ref{thm:reg}, \ref{thm:staticreg}, and \ref{thm:vio}, we can derive the following corollaries regarding the regret and constraint violation bounds. The proofs are omitted for brevity. Here, we define parameters $\delta,\nu,\kappa\ge0$ to represent the time variability of the per-slot optimizers and constraint functions, such that $\Delta_{\mathbf{x}^\star}=\mathcal{O}(T^\delta)$, $\Delta_{\mathbf{g}}=\mathcal{O}(T^\nu)$, and $\tilde{\Delta}_{\mathbf{g}}=\mathcal{O}(T^\kappa)$ \cite{T.Chen}, \cite{X.Cao}. These corollaries provide two sets of performance bounds depending on whether the value of $\delta$ is known.

\setcounter{corollary}{7}
\begin{corollary}\label{cor:dynamic}
\textit{Step-sizes with knowledge of $\delta$:} Let $\alpha=\tau^{-\frac{1}{2}}T^{\frac{1-\delta}{2}}$, $\eta=\beta^2\gamma^2$, and $\gamma=1$ in DTC-OCO. We have
\begin{align}
        \text{RE}_{\text{d}}(T)&=\mathcal{O}\left(\max\left\{\tau^{\frac{1}{2}}T^{\frac{1+\delta}{2}},T^\nu\right\}\right),\\
        \text{RE}_{\text{s}}(T)&=\mathcal{O}\left(\max\left\{\tau^{\frac{1}{2}}T^{\frac{1}{2}},T^\nu\right\}\right),\\
        \text{VO}^c(T)&=\mathcal{O}\left(\max\left\{T^{\frac{1-\delta}{2}},\tau{T}^\kappa\right\}\right).
\end{align}
In particular, if $\tau=\mathcal{O}(1)$, $\delta<1$, $\nu<1$, and $\kappa<1$, both the dynamic and static regrets are sublinear, and the constraint violation is sublinear.
\end{corollary}

\begin{corollary}\label{cor:dynamic2}
\textit{Step-sizes without knowledge of $\delta$}: Let $\alpha=\tau^{-\frac{1}{2}}T^{\frac{1}{2}}$,
$\eta=\beta^2\gamma^2$, and $\gamma=1$ in DTC-OCO. We have
\begin{align}
        \text{RE}_{\text{d}}(T)&=\mathcal{O}\left(\max\left\{\tau^{\frac{1}{2}}T^{\frac{1}{2}+\delta},T^\nu\right\}\right),\\
        \text{RE}_{\text{s}}(T)&=\mathcal{O}\left(\max\left\{\tau^{\frac{1}{2}}T^{\frac{1}{2}},T^\nu\right\}\right),\\
        \text{VO}^c(T)&=\mathcal{O}\left(\max\left\{T^{\frac{1}{2}},\tau{T}^\kappa\right\}\right).\label{eq:vionoT}
\end{align}
\end{corollary}

From Corollaries~\ref{cor:dynamic} and \ref{cor:dynamic2}, a sufficient condition for DTC-OCO to yield sublinear dynamic and static regrets and sublinear constraint violation is that the accumulated variations of the per-slot optimizers and constraints evolve sufficiently slowly. Otherwise, if the system varies too drastically, no online algorithm can track it due to the lack of in-time information \cite{T.Chen}, \cite{X.Cao}, \cite{Gap}. However, in many online applications, the system tends to stabilize over time and sublinear regret and constraint violation can be achieved by DTC-OCO.

\begin{remark}\label{rm:1}
If the feedback delay $\tau$ is unknown, by setting $\alpha=T^{\frac{1-\delta}{2}}$ in DTC-OCO, we can show that $\text{RE}_{\text{d}}(T)=\mathcal{O}(\max\{\tau{T}^{\frac{1+\delta}{2}},T^\nu\})$ and $\text{RE}_{\text{s}}(T)=\mathcal{O}(\max\{\tau{T}^{\frac{1}{2}},T^\nu\})$, both still being sublinear under the same conditions in Corollary~\ref{cor:dynamic}. If both $\tau$ and $\delta$ are unknown, by setting $\alpha=T^{\frac{1}{2}}$ in DTC-OCO, we can show that $\text{RE}_{\text{d}}(T)=\mathcal{O}(\max\{\tau{T}^{\frac{1}{2}+\delta},T^\nu\})$ and $\text{RE}_{\text{s}}(T)=\mathcal{O}(\max\{\tau{T}^{\frac{1}{2}},T^\nu\})$.
\end{remark}

\begin{remark}
With unknown time horizon $T$, the standard doubling trick \cite{OLBK}, \cite{CDC},
\cite{db} can be applied to extend DTC-OCO into one that has similar regret bounds.
\end{remark}

\begin{remark}
The performance analysis in \cite{X.CaoTau} assumes $T$ is large enough such
that $\frac{(1+C)\max\{D,\beta\}^2+2}{\sqrt{\tau T}}+[(5C+1)\max\{D,\beta\}^2+2]\sqrt{\frac{\tau}{T}}\le\sqrt{\frac{1}{3}}$,
and therefore it provides no performance guarantee for a mid-range value of $T$. Furthermore, only a static regret bound is provided in \cite{X.CaoTau}.
In addition, the optimal step-sizes in \cite{X.CaoTau} require knowing the values of $C$ and $D$, which we do not need for DTC-OCO.
\end{remark}

\subsubsection{Special Case of One-Slot Feedback Delay} Since no other algorithm has provided a dynamic regret bound for constrained OCO with multi-slot feedback delay, we next consider the special case of one-slot feedback delay. In the following two remarks, we compare DTC-OCO with \cite{T.Chen} and \cite{X.Cao} under this setting.

\begin{remark}
The dynamic regret and constraint violation bounds achieved by \cite{T.Chen} rely on the key assumption that the slack constant $\epsilon$ is larger than the maximum variation of the constraints, \ie $\epsilon>\max_{t\in[1,T]}\left\{\max_{\mathbf{x}\in\mathcal{X}_0}\left\{ \Vert\mathbf{g}_t(\mathbf{x})-\mathbf{g}_{t-1}(\mathbf{x})\Vert_2\right\}\right\}$, which may be difficult to satisfy in general. In contrast, DTC-OCO only assumes $\epsilon>0$ as indicated in (\ref{eq:epsilon}). Furthermore, the optimal step-sizes used in \cite{T.Chen} require knowledge of the accumulated variation of the constraint function sequence~$\kappa$, while DTC-OCO only needs an upper bound on its gradient $\beta$, which is much easier to acquire than $\kappa$. When $\delta$ is unknown, \cite{T.Chen} achieves $\mathcal{O}(\max\{T^{\frac{1}{3}+\delta},T^{\frac{1}{3}+\kappa},T^{\frac{2}{3}}\})$
dynamic regret and $\mathcal{O}(T^{\frac{2}{3}})$ constraint violation,
both being at least $\mathcal{O}(T^{\frac{2}{3}})$. In contrast, the performance bounds of DTC-OCO decreases smoothly to $\mathcal{O}(T^{\frac{1}{2}})$ if the system variation is sufficiently small.
\end{remark}

\begin{remark}
To achieve sublinear dynamic regret and constraint violation, \cite{X.Cao} relies on additional assumptions that the accumulated variation of the convex loss functions $\{f_t(\mathbf{x})\}$ is sublinear regardless of the trajectory of the online decision sequence, \ie $\sum_{t=1}^{T}\max_{\mathbf{x}\in\mathcal{X}_0}\left\{\left\Vert{f}_{t}(\mathbf{x})-f_{t-1}(\mathbf{x})\right\Vert_2\right\}=\mathbf{o}(T)$,
and the accumulated variation of the optimal dual points $\{\lambda_t^\star\}$
of the optimization problem $\textbf{P1}$ is sublinear, \ie $\sum_{t=1}^T\Vert\lambda_{t+1}^\star-\lambda_t^\star\Vert_2=\mathbf{o}(T)$. Neither assumption is required for DTC-OCO. Furthermore, \cite{X.Cao} requires complete information feedback of the convex loss function $f_t(\mathbf{x})$, and therefore has the gradient information at
any point and can even penalize the loss function directly. In contrast, DTC-OCO requires only the gradient information of the convex loss function at the online decision points $\nabla f_t(\mathbf{x}_t)$.
\end{remark}

\subsubsection{Special Case of Time-invariant Constraints}

When the constraints are time-invariant, the following corollary suggests that DTC-OCO recovers the known best $\mathcal{O}(\tau^{\frac{1}{2}}T^{\frac{1}{2}})$ static regret for unconstrained OCO with multi-slot delay \cite{Slow}, and its $\mathcal{O}(\tau)$ constraint violation recovers the known best $\mathcal{O}(1)$ constraint violation under the standard OCO setting of one-slot feedback delay \cite{LTC-HY} as a special case.

\begin{corollary}\label{cor:inv}
\textit{Time-invariant constraints:} Let $\alpha=\gamma^2=\tau^{-\frac{1}{2}}T^{\frac{1-\delta}{2}}$ and $\eta=\beta^2\gamma^2$ in DTC-OCO. We have $\text{RE}_{\text{d}}(T)=\mathcal{O}(\tau^{\frac{1}{2}}T^{\frac{1+\delta}{2}})$,
$\text{RE}_{\text{s}}(T)=\mathcal{O}(\tau^{\frac{1}{2}}T^{\frac{1}{2}})$, and $\text{VO}^c(T)=\mathcal{O}(\tau )$.

In particular, if $\tau=\mathcal{O}(1)$ and $\delta<1$, both the dynamic and static regrets are sublinear,
and the constraint violation is upper bounded by a constant.\end{corollary}

\begin{remark}
The constraint violation bounds in \cite{T.Chen} and \cite{X.CaoTau} are no less than $\mathcal{O}(T^\frac{2}{3})$ and $\mathcal{O}(T^\frac{3}{4})$, respectively, even if the constraint function is time-invariant. In contrast, when $\tilde{\Delta}_{\mathbf{g}}$ is small, the constraint violation $\text{VO}^c(T)$ yielded by DTC-OCO decreases smoothly to $\mathcal{O}(T^\frac{1}{2})$. Particularly, if the constraint function is time-invariant, \ie $\tilde{\Delta}_{\mathbf{g}}=0$, DTC-OCO provides $\mathcal{O}(\tau)$ constraint violation.
\end{remark}

\section{Application to Network Resource Allocation}
\label{Sec:Application to Network Resource Allocation}

We apply DTC-OCO to a general network resource allocation problem \cite{T.Chen}, \cite{X.Cao}, \cite{Neely}, \cite{RAWN}, \cite{T.ChenLy}. We present numerical results to demonstrate the performance advantage of DTC-OCO over the known best alternative from \cite{X.CaoTau}.

\subsection{Online Network Resource Allocation}
\label{Subsec:Online Network Resource Allocation}

\begin{figure}[t]
\centering
\includegraphics[width=.75\linewidth,trim= 0 0 0 0,clip]{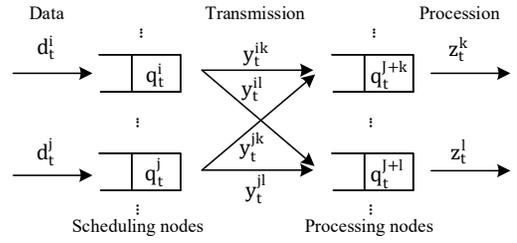}
\vspace{-1mm}
\caption {An illustration of general online network resource allocation.}
\label{fig:NRA}
\vspace{-4mm}
\end{figure}

Fig.~\ref{fig:NRA} shows a general network consisting of $J$ scheduling nodes and $K$ processing nodes. For example, in a wired or wireless network, the scheduling nodes may be relays, and the processing nodes may be sink nodes. In a cloud computing network, the scheduling nodes may be mappers, and the processing nodes may be computing servers.

At each time slot $t$, the amount of data arriving at scheduling node $j$ is denoted by $d_t^j$, and we define an extended data arrival vector  denoted by $\mathbf{d}_t=[d_t^1,\dots,d_t^J,\mathbf{0}_{1\times{K}}]^T$. A central controller decides the transmission rate $y_t^{jk}$ of the link $(j,k)$ connecting scheduling node $j$ and processing node $k$, and the processing rate $z_t^k$ at processing node $k$. In compact form, the decision vector at time $t$ is $\mathbf{x}_t=[y_t^{11},\dots,y_t^{JK},z_t^1,\dots,z_t^K]^T$. Denote the maximum data transmission rate of link $(j,k)$ by $y_{\text{max}}^{jk}$, and the maximum data processing rate of processing node $k$ by $z_{\text{max}}^k$. The data rate limits are compactly expressed in the convex set as $\mathcal{X}_0\triangleq\{\mathbf{x}|\mathbf{0}\preceq\mathbf{x}\preceq\mathbf{x}_{\text{max}}\}$, where $\mathbf{x}_{\text{max}}=[y_{\text{max}}^{11},\dots,y_{\text{max}}^{JK},z_{\text{max}}^{1},\dots,z_{\text{max}}^{K}]^T$ is the maximum data rate vector. Each scheduling node $j$ and processing node $k$ has a local data queue backlog at time $t$ denoted by $q_t^j$ and $q_t^{J+k}$, respectively. The queue backlog vector is $\mathbf{q}_t=[q_t^1,\dots{q}_t^J,q_t^{J+1},\dots,q_t^{J+K}]^T$, and we can express its update as $\mathbf{q}_{t+1}=[\mathbf{q}_t+\mathbf{C}\mathbf{x}_t+\mathbf{d}_t]^+$, where $\mathbf{C}\in\mathbb{R}^{(J+K)\times(JK+K)}$ represents the network topology and is given by
\begin{align*}
\mathbf{C}=
\left[\begin{array}{c | c}
\blkdiag\{-\mathbf{1}_{1\times{K}},\dots,-\mathbf{1}_{1\times{K}}\} & \mathbf{0}_{J\times{K}}\\\hline
\mathbf{I}_{K\times{K}},\dots, \mathbf{I}_{K\times{K}} & -\mathbf{I}_{K\times{K}}
\end{array}\right]_{}.
\end{align*}

The goal for a network controller is to minimize the network cost while controlling the long-term averaged data outgoing rate to be no less than the incoming data rate for queue stability. However, since the controller can receive only delayed feedback on system parameters $\mathbf{d}_t$, $\mathbf{q}_t$, and $f_t(\mathbf{x})$ over time, it must employ an online solution. This online network resource allocation problem is a special case of the OCO problem $\textbf{P1}$, with the convex set $\mathcal{X}_0$ define above, for any convex loss function $f_t(\mathbf{x})$, and the convex constraint function $\mathbf{g}_t(\mathbf{x})\triangleq\mathbf{C}\mathbf{x}+\mathbf{d}_t$, which represents the difference between incoming and outgoing data at time $t$. Note that achieving sublinear constraint violation, \ie $\lim_{T\to\infty}\frac{1}{T}\sum_{t=1}^T\mathbf{g}_t(\mathbf{x}_t)\to\mathbf{0}$, is equivalent to queue stability.

A special case of this problem where there is no delay in system information has been addressed with Lyapunov optimization techniques \cite{Neely}, \cite{RAWN}, \cite{T.ChenLy}. Furthermore, solutions under the standard OCO setting with one-slot feedback delay are given in \cite{T.Chen} and \cite{X.Cao}. However, due to the intermittence and crowdedness of the communication links between the nodes and central controller, it is likely that the central controller experiences multi-slot feedback delay of the system parameters. The proposed DTC-OCO algorithm provides a suitable online solution to this problem.

\subsection{Numerical Performance Evaluation}
\label{Subsec:Simulation Results}

When applying DTC-OCO to the above online network resource allocation problem, the sublinear regret bounds derived in Section~\ref{Sec:Performance Bounds} imply that it will provide an efficient solution. Furthermore, the sublinear constraint violation bound guarantees queue stability. In this section, we further study the numerical performance of DTC-OCO in a practical setting, compare its performance with \cite{X.CaoTau}, and verify the benefit of the proposed double regularization.

As a tangible example, we consider a mobile cloud computing system consisting of $J=10$ scheduling nodes, and $K=10$ processing nodes. Following the typical long-term evolution (LTE) specifications \cite{LTEP}, we set the noise power spectral density $N_0=-174$~$\text{dBm}/\text{Hz}$, noise figure $N_F=10$~dB, and channel bandwidth $B_W=10$~MHz as default system parameters. We set the time slot duration to be $1$~ms and assume $\mathbf{d}_t$ kB of data arrive at each time $t$. The maximum data transmission and processing rates, in $\text{MB}/\text{s}$, are randomly uniformly distributed as $y_{\text{max}}^{jk}\sim\mathcal{U}(10,100)$ and $z_{\text{max}}^k\sim\mathcal{U}(100,250)$, respectively. According to the Shannon bound, we consider the transmission power of each link $(j,k)$ as exponential w.r.t. its transmission rate $y^{jk}$ given by $\frac{\sigma_n^2}{L_t^{jk}}(2^\frac{y^{jk}}{B_W}-1)$,
where $\sigma_n^2[\text{dBm}]=N_0B_W+N_F$ is the noise power and $L_t^{jk}$ represents
the integrated impact of path-loss, interference, and capacity gap, and it is time-varying due to the fluctuation of wireless channels. We assume each processing node $k$ follows a quadratic power-frequency relationship given by $\theta(\xi_t^k z^k)^2$, where $\theta=120~\text{W}/(\text{GHz})^2$ \cite{ProCost} and $\xi_t^k$, in cycles per byte, depends on the computational complexity of the computing tasks processed by node $k$ at time $t$ \cite{Price}, \cite{Sundar18} and thus is time varying \cite{TVPrice}. We consider both the data transmission and processing power by defining the following cost function at time $t$:
\begin{align*}
        f_t(\mathbf{x})\triangleq\sum_{j\in\mathcal{J}}\sum_{k\in\mathcal{K}}\frac{\sigma_n^2}{L_t^{jk}}(2^\frac{y^{jk}}{B_W}-1)+\sum_{k\in\mathcal{K}}\theta(\xi_t^k{z}^k)^2.
\end{align*}
We remark here that our proposed OCO solution can be applied to more general cost functions, as long as they are convex w.r.t. the decision variables $\mathbf{x}$.

\begin{figure}[!t]
\centering
\vspace{-2mm}
\subfloat[I.i.d. parameters.\label{fig:iid}]
{\includegraphics[width=.75\linewidth,trim= 10 45 10 00,clip]{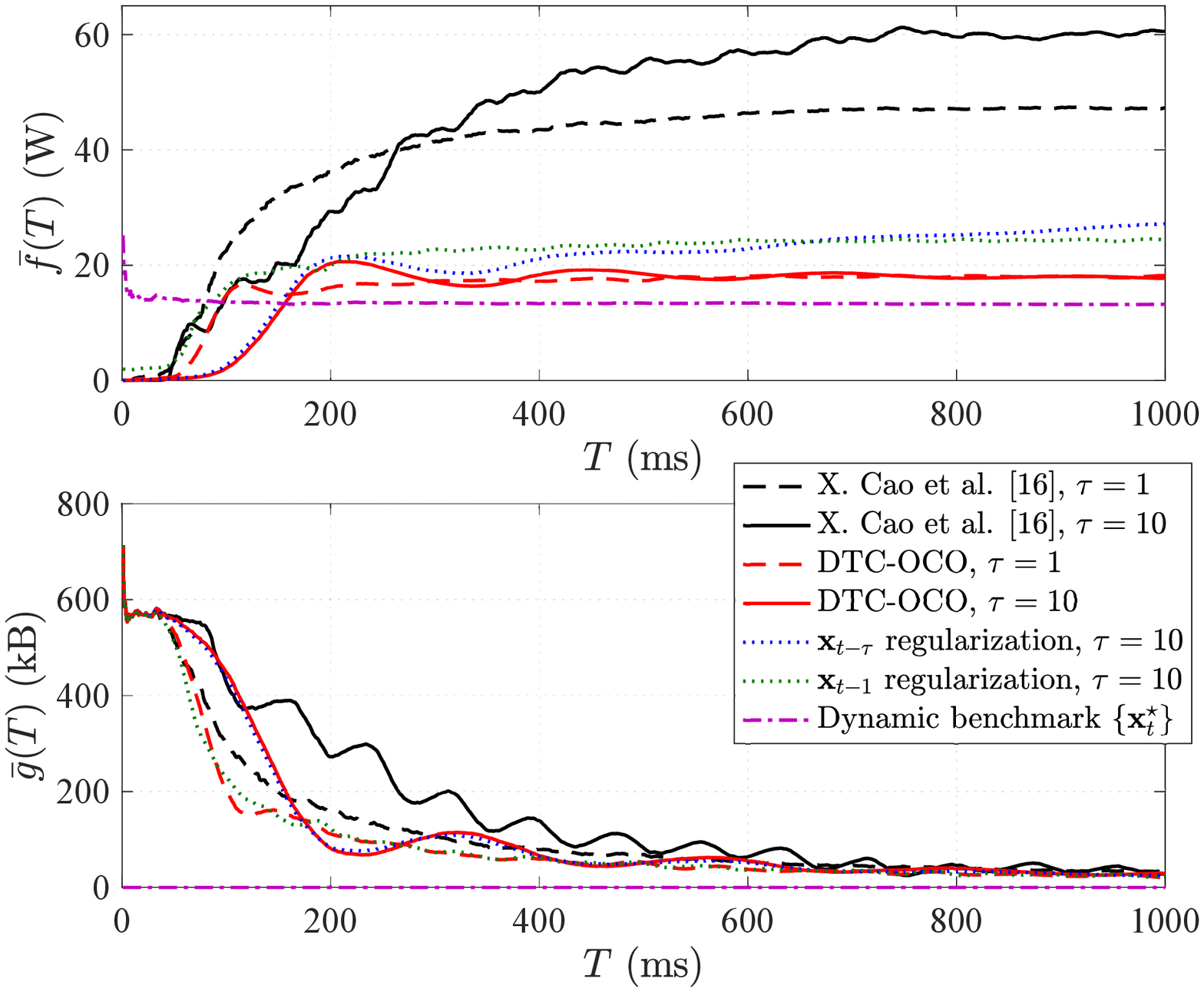}}
\vspace{-0mm}
\subfloat[Noisy periodic parameters.\label{fig:periodic}]
{\includegraphics[width=.75\linewidth,trim= 10 45 10 00,clip]{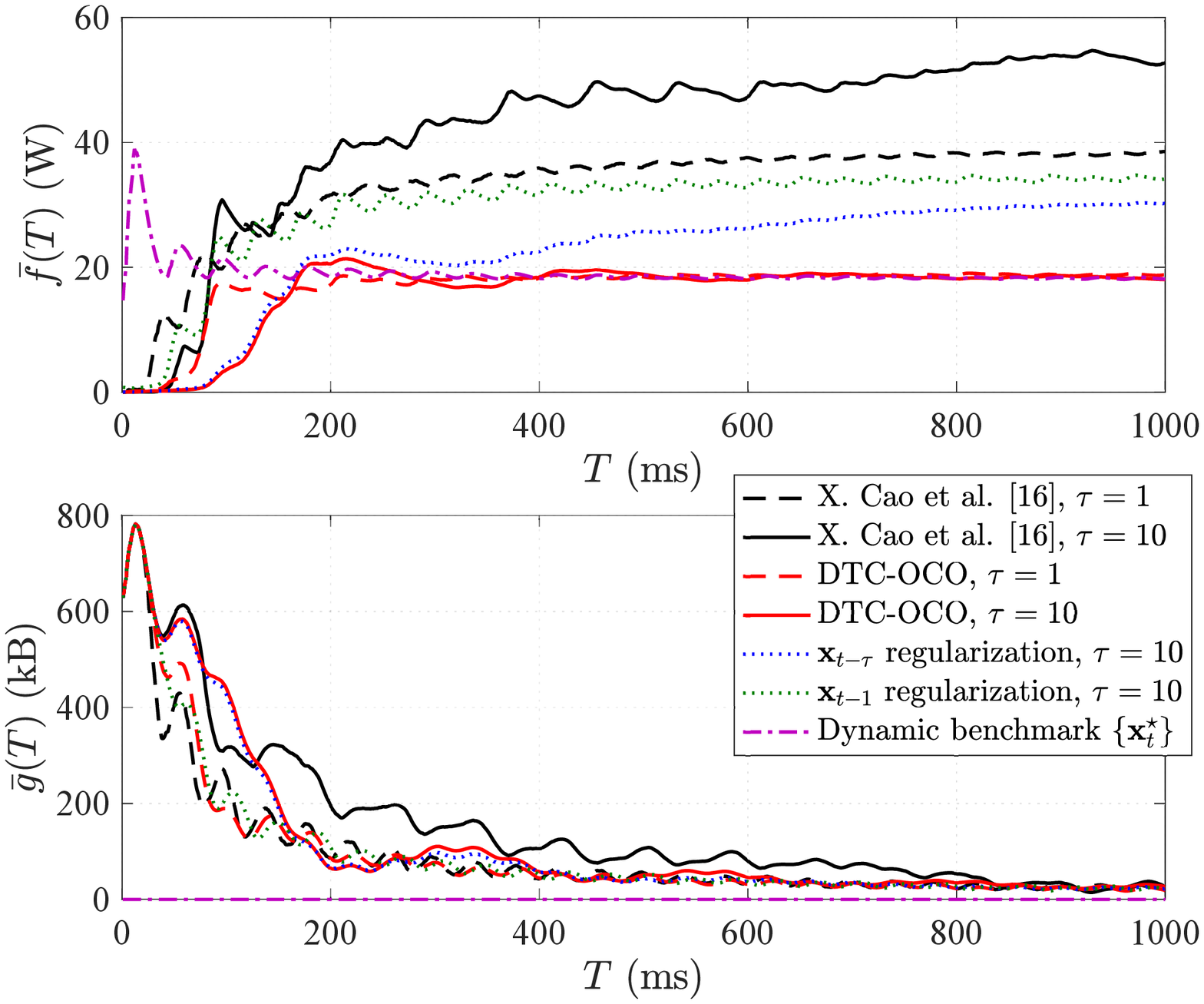}}
\vspace{-1mm}
\caption{$\bar{f}(T)$ and $\bar{g}(T)$ vs. $T$ under different $\tau$.}\label{fig:Comp}
\vspace{-4mm}
\end{figure}

For performance comparison, we consider the online algorithm proposed in \cite{X.CaoTau}, which is the known best time-varying constrained OCO algorithm that accommodates multi-slot feedback delay. We also consider the dynamic benchmark in (\ref{eq:xtstar}). To verify the performance advantage brought by the double regularization in DTC-OCO, we also study two simplified versions of DTC-OCO with single regularization on $\mathbf{x}_{t-\tau}$ and $\mathbf{x}_{t-1}$, respectively.

We assume both $\tau$ and $\delta$ are unknown, and thus set $\alpha=T^{\frac{1}{2}}$, $\eta=\Vert\mathbf{C}\Vert_2^2$, and $\gamma=1$ in DTC-OCO (see Remark \ref{rm:1}). Our performance metrics are the time-averaged network cost $\bar{f}(T)\triangleq\frac{1}{T}\sum_{t=1}^Tf_t(\mathbf{x}_t)$
and the time-averaged constraint violation $\bar{g}(T)\triangleq\frac{1}{T}\sum_{t=1}^T\mathbf{1}^T\mathbf{g}_t(\mathbf{x}_t)$. For fair comparison on $\bar{f}(T)$, the step-sizes in \cite{X.CaoTau} are selected such that the algorithm has a similar steady state value of $\bar{g}(T)$ as DTC-OCO. The time-varying system parameters $\{d_t^j\}$, $\{L_t^{jk}\}$, $\{\xi_t^k\}$ are modeled by the following two cases.

\textit{1) I.i.d. parameters.}
The system parameters are i.i.d.: $d_t^j\sim\mathcal{U}(10,100)$,
$L_t^{jk}[\text{dB}]\sim\mathcal{U}(-126,-120)$, and $\xi_t^k\sim\mathcal{U}(1,3)$.

\textit{2) Noisy periodic parameters.}
The system parameters vary periodically with noise: $d_t^j=30\sin\left(\frac{\pi{t}}{20}\right)+n_t^{j,d}$, $L_t^{jk}[\text{dB}]=-120-3\sin\left(\frac{\pi{t}}{20}\right)-n_t^{jk,L}$, and $\xi_t^k=0.5\sin(\frac{\pi{t}}{20})+n_t^{k,\xi}$, where $n_t^{j,d}\sim\mathcal{U}(40,70)$, $n_t^{jk,L}\sim\mathcal{U}(6,9)$, and $n_t^{k,\xi}\sim\mathcal{U}(1,3)$.

Fig.~\ref{fig:Comp} shows $\bar{f}(T)$ and $\bar{g}(T)$ versus $T$ with different values of $\tau$ for the above two cases. We observe that the network cost of DTC-OCO can approach that of the dynamic benchmark in (\ref{eq:xtstar}), indicating that sublinear dynamic regret is achieved.  Compared with \cite{X.CaoTau}, DTC-OCO achieves lower network cost and is much more tolerant to the feedback delay $\tau$. The reason for this are two fold: first, DTC-OCO penalizes the constraint function $\mathbf{g}_t(\mathbf{x})$ directly instead of its first-order approximation, which improves the control on constraint violation; second, the double regularization in DTC-OCO allows the online decision $\mathbf{x}_t$ to be updated from $\mathbf{x}_{t-\tau}$ and $\mathbf{x}_{t-1}$, which improves the efficacy of gradient descent for loss minimization. Comparison with the $\mathbf{x}_{t-\tau}$ and $\mathbf{x}_{t-1}$ single-regularization approaches further suggests that the proposed method of double regularization is essential to the performance advantage of DTC-OCO.

\section{Conclusions}
\label{Sec:Conclusions}

This paper considers OCO with multi-slot feedback delay and short-term and long-term constraints. We propose an efficient algorithm DTC-OCO, which uses a novel constraint penalty with double regularization to handle the asynchrony between information feedback and decision updates. Our analysis considers the impact of multi-slot feedback delay and the double regularization structure on the performance guarantees of DTC-OCO, to show sublinear dynamic and static regrets and sublinear constraint violation under mild conditions. We apply DTC-OCO to a general online network resource allocation problem, using mobile cloud computing for numerical example. Simulation results demonstrate the superior delay tolerance and substantial performance advantage of DTC-OCO over the known best alternative under different scenarios.
\balance

\bibliographystyle{IEEEtran}
\bibliography{INFOCOM2021}

\end{document}